\documentclass[twocolumn]{aastex631}

\newcommand{\degrees}{$^{\circ}$}

\begin{document}

\title{Warm Jupiters Beyond the Tidal Synchronization Limit May Exhibit a Wide Range of Secondary Eclipse Depths}

\correspondingauthor{Emily Rauscher}
\email{erausche@umich.edu}

\author[0000-0003-3963-9672]{Emily Rauscher}
\affiliation{Department of Astronomy, University of Michigan, 1085 S. University Ave., Ann Arbor, MI 48109, USA}

\author[0000-0001-6129-5699]{Nicolas B. Cowan}
\affiliation{Department of Physics, McGill University, 3600 rue University, Montréal, QC, H3A 2T8, CAN}
\affiliation{McGill Space Institute, 3550 rue University, Montréal, QC, H3A 2A7, CAN}

\author[0000-0002-0296-3826]{Rodrigo Luger}
\affiliation{Center for Computational Astrophysics, Flatiron Institute, New York, NY, USA}

\begin{abstract}

With JWST we can now characterize the atmospheres of planets on longer orbital planets, but this moves us into a regime where we cannot assume that tidal forces from the star have eroded planets' obliquities and synchronized their rotation rates. These rotation vectors may be tracers of formation and evolution histories and also enable a range of atmospheric circulation states. Here we delineate the orbital space over which tidal synchronization and alignment assumptions may no longer apply and present three-dimensional atmospheric models of a hypothetical warm Jupiter over a range of rotation rates and obliquities. We simulate the secondary eclipses of this planet for different possible viewing orientations and times during its orbital, seasonal cycle. We find that the eclipse depth can be strongly influenced by rotation rate and obliquity through the timing of the eclipse relative to the planet's seasonal cycle, and advise caution in attempting to derive properties such as albedo or day-night transport from this measurement. We predict that if warm Jupiters beyond the tidal limit have intrinsic diversity in their rotation vectors, then it will manifest itself as dispersion in their secondary eclipse depths. We explore eclipse mapping as a way to uniquely constrain the rotation vector of warm Jupiters but find that the associated signals are likely at the edge of JWST performance. Nevertheless, as JWST begins to measure the secondary eclipses of longer orbital period planets, we should expect to observe the consequences of a wider range of rotation states and circulation patterns.

\end{abstract}


\section{Introduction} \label{sec:intro}

One of the exciting developments in exoplanet characterization, especially as we enter the new JWST era, is the expansion of feasible observational targets to include ones beyond the hot Jupiter population. By observing planets that are smaller and that orbit farther from their host stars, we can expand our understanding of planetary physics into new regimes, bridging the expanse toward Solar System conditions. In this paper we focus on the characterization of gaseous planets on longer orbital periods, namely the ``warm" Jupiter population. We expect to encounter new chemical regimes with these slightly cooler planets, as different molecular species are expected to be dominant in chemical equilibrium; for example, methane should take over as the primary carbon-bearing molecule, although current observations suggest many warm gaseous planets may be depleted in this species \citep{Stevenson2010,Madhusudhan2011b,Kreidberg2018b,Benneke2019}. In addition, these planets have been identified as important targets for understanding planets' global compositions, as their radii are not subject to anomalous inflation like many hot Jupiters \citep{Thorngren2016}. 

The formation pathway for warm Jupiter planets is directly related to the question of hot Jupiter formation and currently a murky picture exists where different pieces of empirical evidence point toward disparate formation mechanisms, perhaps indicating that multiple pathways are at work \citep[see review by][]{Dawson2018}. The rotation state of warm Jupiters will be set by their formation and then subsequently modified by their orbital evolution. The rotation speed of gas giants is inherited from the angular momentum of the gas from which they form and thought to be regulated by magnetic interaction with the circumplanetary disk \citep{Takata1996,Batygin2018,Ginzburg2020}. The initial rotational obliquity of a planet could be misaligned with its orbital axis through formation conditions \citep{Tremaine1991,Millholland2019b} and excited (or damped) through various processes after formation \citep{Ward2004,Ward2006,Li2020,Rogoszinski2020,Rogoszinski2021,Saillenfest2021a}, including processes related to any migration the planet may undergo \citep{Millholland2019a,Hong2021}. Empirical determination of the obliquities of planetary mass companions on wide orbits are beginning to constrain their formation and evolution \citep{Bryan2020a,Bryan2021}.

As more bright warm Jupiter planets are identified and their properties characterized \citep[e.g.,][]{Dong2021}, we can hope to gain a clearer understanding of planet formation overall. In particular, warm Jupiters may maintain non-zero rotational obliquities (i.e., axial tilts) and faster rotation rates than hot Jupiters, and this could inform their histories of formation and subsequent evolution. Meanwhile, non-zero obliquities can induce seasonal variations in the atmospheres of these planets, introducing a wider range of possible states, potentially confusing the interpretation of characterization measurements. Here we consider the linked questions of: 1) how the \textit{a priori} unknown rotation state of a warm Jupiter could impact its atmospheric structure, with implications for its secondary eclipse and 2) whether its rotation rate and obliquity could be empirically constrained from atmospheric characterization measurements. This work uses a similar set-up as \citet{Rauscher2017}, but expands to include both different obliquities and a range of rotation rates. 

We discuss estimates of tidal alignment and circularization, with a focus on warm Jupiters, in Section \ref{sec:vector_expectations}. In Section \ref{sec:models} we present our three-dimensional atmospheric model of a hypothetical warm Jupiter, for various rotation rates and obliquities. We present simulated secondary eclipses from those models in Section \ref{sec:eclipses} and consider whether we could retrieve spatial information from these by using JWST eclipse mapping in Section \ref{sec:mapping}. We discuss some caveats to this work in Section \ref{sec:discussion} and summarize our findings in Section \ref{sec:conclusions}.

\section{Tidal Synchronization and Alignment of Warm Jupiter Rotation Vectors} \label{sec:vector_expectations}

Hot Jupiter planets are so close to their host stars that we expect tidal forces from the star to have put them into a rotation state that is synchronous (the rotation period equals the orbital period) and aligned (the rotation axis is parallel to the orbital axis). While the rotation and alignment of hot Jupiters has yet to be unambiguously empirically constrained \citep{Adams2019,Flowers2019,Beltz2021}, our expectation for the longer orbital period warm Jupiters should be that they have a range of rotation rates and obliquities. 
We can estimate the timescales for synchronization and alignment \citep[which are within a factor of two of each other][]{Peale1999,Fabrycky2007} using the following expression from \citet{Guillot1996}, re-written using Kepler's 3rd Law to to show the dependence on orbital period ($P$) instead of semi-major axis:
\begin{eqnarray}
   \tau_{\mathrm{tidal}}   & \sim & \left(\frac{G}{16\pi^4}\right) Q \omega_p \left(\frac{M_p}{R_p^3}\right) P^4 \\  \label{eqn:tidal}
    \tau_{\mathrm{tidal}}\ (\mathrm{Gyr})  & \sim & 0.067 \left(\frac{Q}{10^{5}}\right) \left(\frac{\omega_p M_p R_J^3}{\omega_J M_J R_p^3}\right) \left(\frac{P}{10\ \mathrm{days}}\right)^4 \nonumber
\end{eqnarray}
where $M_p$, $R_p$, and $\omega_p$ are the mass, radius, and rotation rate of the planet, the subscript $J$ references Jupiter's values for these quantities, and $Q$ is the planet's tidal dissipation factor, which simplifies the complexities of tidal interactions into a single, uncertain parameter.

\begin{figure*}
    \centering
    \includegraphics[width=0.95\textwidth]{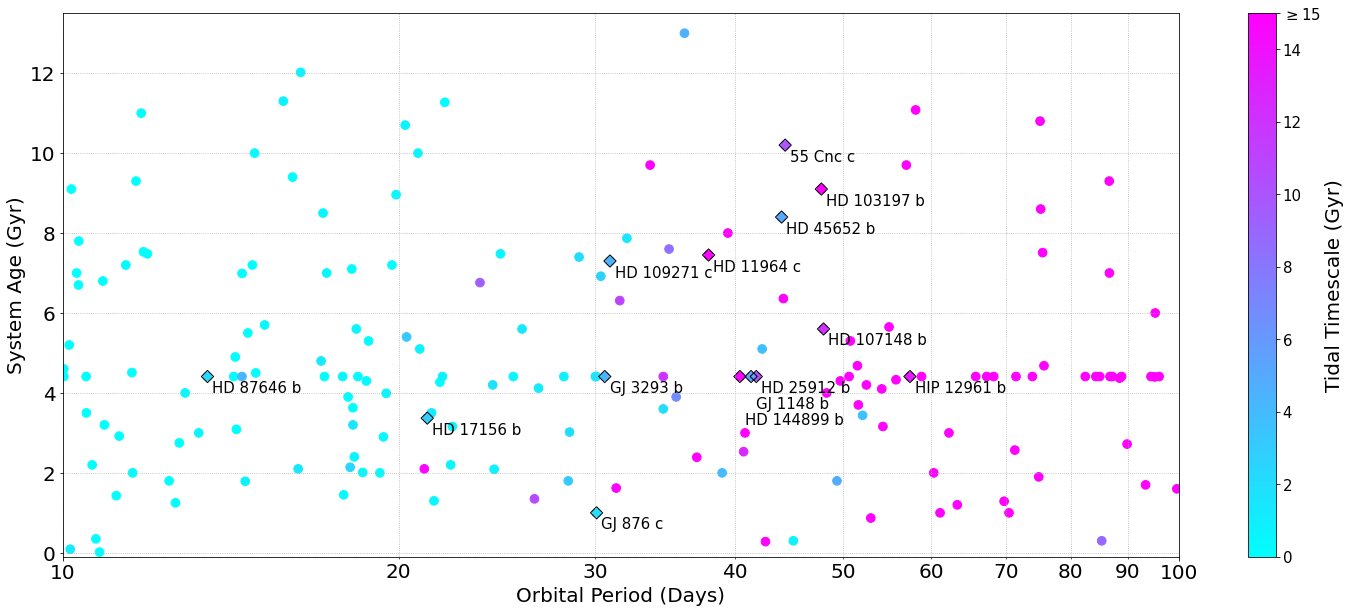}
    \caption{Estimated timescales for tidal synchronization and alignment (from Equation \ref{eqn:tidal}) versus orbital period and system age for gaseous exoplanets ($R_p>0.4R_J$) with measured masses. No errors are shown, but generally the orbital periods are very well known while the system ages are not (and here are set to 4.14 Gyr if not listed). Particularly bright planets that are near the boundary where tidal timescales equal system ages ($0.25 <$ age/$\tau_{\mathrm{tidal}}<2$) are labeled.}
    \label{fig:knownsystems}
\end{figure*}

In Figure \ref{fig:knownsystems} we plot estimates for tidal synchronization and alignment timescales, compared to host star ages and over a range of orbital periods, for known exoplanets\footnote{From the NASA Exoplanet \cite{exoplanetarchive}
, accessed November 29, 2022} with measured masses and radii (and $R_p>0.4 R_J$, to isolate gas giants). In all cases we assume an initial rotation rate and tidal $Q$ appropriate for Jupiter \citep[$\omega_J=1.7 \times 10^{-4}$ s$^{-1}$, $Q=10^5$;][]{Guillot1996}. Age estimates of the host stars can be highly uncertain and in some cases are missing; for the stars without a listed age, we use the mean value for this population (4.14 Gyr). Based on these estimates, planets with orbital periods longer than $\sim$30 days have tidal timescales longer than the system ages and so should not have had time for the planets' rotation vectors to synchronize and align. We highlight planets near the tidal boundary that are particularly bright, with host star K magnitudes less than or equal to WASP-18 \citep[$K=8.131$, the secondary eclipse target for the JWST Transiting Planet ERS Program;][]{Bean2018}. These are promising prospects for measuring secondary eclipses of planets past the tidal boundary with JWST, as we will further elaborate on in Section \ref{sec:eclipses}.

\section{Three-Dimensional Models of a Hypothetical Warm Jupiter} \label{sec:models}

Having the rotation vector of a planet \textit{a priori} unknown (or un-assumed) expands the possible circulation states for its atmosphere. Previous models of warm Jupiters have explored the role of rotation rate \citep{Showman2015}, obliquity \citep{May2016,Rauscher2017}, and the interaction between these two parameters and additionally any orbital eccentricity \citep{Ohno2019a}. In order to explore a wide parameter space, \citet{Ohno2019a} used a simplified ``shallow water" model with Newtonian cooling to simulate the atmospheric dynamics for gaseous planets with obliquities from 0-180\degrees, eccentricities of 0, 0.3, and 0.5, and three radiative timescales, which defined different forcing regimes (similar to hot Jupiters, warm Jupiters, and cold Jupiters). The authors identified distinct dynamical regimes, determined by obliquity and radiative timescale, while relatively insensitive to eccentricity. For planets with radiative timescales less than the rotation period (such as synchronously rotating hot or warm Jupiters), the atmosphere is dominated by a day-night contrast and a shifted hot spot. For planets where the radiative timescale is less than the orbital period but longer than the rotation period (such as quickly rotating warm Jupiters), the diurnal mean forcing is dominant and the equator is hotter than the poles for obliquities less than $\sim18$\degrees, while higher obliquities result in seasonal cycles, with the hottest region of the atmosphere moving back and forth between the poles. These results were then used in \citet{Ohno2019b} to calculate simulated phase curves from the models; the authors found that the planet's eccentricity strongly influences the phase at which peak flux may be measured, as does the viewing orientation for planets with significant obliquity \citep[matching the results from][]{Rauscher2017}. They also created an analytic theory to describe the phase curve behavior.

Here we focus on just the secondary eclipses of warm Jupiters, which are less observationally expensive than full-orbit phase curves (especially for orbital periods $\ge$10 days) and can still constrain the atmospheric state and circulation pattern of these planets. For this study we construct a hypothetical warm Jupiter with the mass and radius of Jupiter, an orbital period of 10 days, and a Solar twin as its host. With these properties, the planet would only have a tidal timescale comparable to its system age in a fairly young system ($\lesssim$1 Gyr). Planets on slightly longer orbital periods have much longer tidal timescales and so are less likely to be synchronized and aligned in more mature systems. The parameters for our hypothetical warm Jupiter still place it within the range of known systems and we choose these somewhat optimistically in order to search for observable features: by being closer to the star, the planet will be overall hotter, brighter, and have larger temperature differences. We also choose a circular orbit for our warm Jupiter, as our goal is to investigate the observable implications of the rotation rate and obliquity. We note that a warm Jupiter eccentricity could be constrained from high quality radial velocity data, providing a possible empirical constraint unavailable to the rotation vector.

\subsection{Numerical Set-Up}

To create a set of simulated atmospheric states for this hypothetical warm Jupiter, we use the three-dimensional atmospheric circulation model RM-GCM \citep{Menou2009,Rauscher2010,Rauscher2012b}, which solves a standard, simplified form of the full set of fluid equations, the ``primitive equations of meteorology". Here we use the version that is coupled with a two-stream double-gray radiative transfer routine, updated as described in \citet{Roman2017} to use the numerical scheme from \citet{Toon}. Table \ref{tab:params} lists the planetary parameters for our hypothetical warm Jupiter. The irradiation level assumes a Sun-like host and corresponds to an equilibrium temperature of 880 K. We use the same set of obliquities sampled in \citet{Rauscher2017}, which only presented models for a warm Jupiter rotating as quickly as Jupiter (24.184 $\omega_\mathrm{orb}$). Here we consider a range of possible rotation rates, using a synchronous value and Jupiter's rotation rate as reasonable bookends on possible values. Based on some preliminary zero-obliquity models at more finely sampled rotation rates, we decided to run the full suite using the values shown in Table \ref{tab:params} to capture the full range of zonal jet patterns (equatorial vs. high-latitude), photospheric temperature patterns (day-night vs. azimuthally symmetric), and equator-to-pole temperature gradients. Since the radiative transfer routine of RM-GCM was updated since it was used in \citet{Rauscher2017}, we have recomputed the most quickly rotating models for this work and find no significant differences in the atmospheric structures.

\begin{deluxetable}{lcc}
\tablecaption{Warm Jupiter Model Parameters}
\tablehead{\colhead{Parameter} & \colhead{Value} & \colhead{Units}} 
\startdata
Planetary radius, $R_p$             & $6.986\times10^7$     & m \\
Surface gravity, $g$                & 26                    & m s$^{-2}$ \\
Orbital period, $P_\mathrm{orb}$    & 10                    & days \\
Orbital revolution rate, $\omega_\mathrm{orb}$ 
                                    & $7.2722\times10^{-6}$ & s$^{-1}$ \\
Semi-major axis, $a$                & 0.090867              & AU \\
Eccentricity                        & 0                     & -- \\
Stellar flux at substellar point    & $1.36\times10^5$      & W m$^{-2}$ \\
Albedo                              & 0                     & -- \\
Equilibrium temperature             & 880                   & K \\
Internal heat flux                  & 5.7                   & W m$^{-2}$ \\
Optical abs.\ coefficient, $\kappa_{\mathrm{opt}}$ 
                                    & $2.6\times10^{-3}$    & cm$^2$ g$^{-1}$ \\
Infrared abs.\ coefficient, $\kappa_{\mathrm{IR}}$ 
                                    & $5.2\times10^{-2}$    & cm$^2$ g$^{-1}$ \\
\hline
Rotation rate       & 1/[1, 2, 4, 12, 24.184]   & $\omega_\mathrm{Jup}$ \\
Obliquity           & [0, 10, 30, 60, 90]       & \degrees \\
\hline
Orbital inclination & 88.5      & \degrees \\
Impact parameter    & 0.511     & -- \\
\enddata
\tablecomments{The planet's radius and gravity values are equal to those of Jupiter. The rotation rates are given relative to that of Jupiter, $\omega_\mathrm{Jup}=1.7587 \times 10^{-4}$ s$^{-1}$, with $\omega_\mathrm{Jup}/24.184=\omega_\mathrm{orb}$ being the synchronously rotating model. The stellar parameters are set equal to those of the Sun.} \label{tab:params}
\end{deluxetable}

For the most quickly rotating models, we determined in \citet{Rauscher2017} that it was appropriate to use a diurnal averaging for the stellar irradiation pattern (i.e., instead of tracking the illuminated hemisphere, the incident flux at the top of the atmosphere varies only with latitude, averaged over one planetary rotation). The physical basis for this is roughly that the timescale over which the gas heats and cools is longer than the rotation period \citep[see][for the more exactly defined physical requirement]{Showman2015}. Here we again use this diurnal averaging for the Jupiter-rotation models, but for all of the other models we explicitly track the movement of the stellar irradiation pattern throughout the simulation. We used a timestep between radiative transfer calculations (for all models) such that the intervening movement of the substellar point was always less than a resolution element.

The horizontal resolution for all models is T42 (a truncation of spectral modes beyond 42), corresponding to $\sim$2.8\degrees~at the equator. As in \citet{Rauscher2017}, this was deemed to sufficiently resolve dynamical scales for the most quickly rotating model (where the scales are the smallest), using an estimate of the Rossby deformation radius to be $\sim$8\degrees. Our modeled atmosphere extends from 100 bar to 1 mbar, with 30 vertical levels evenly spaced in $\log$ pressure. We use hyperdissipation to remove noise at the smallest scales of the simulation, acting as an eighth-order operator on the wind and temperature fields, with a timescale (for all models) of about 720 seconds. All simulations used a timestep of $\sim$100 seconds, with radiative fluxes updated every 5 timesteps. We started each simulation at rest (zero winds), with a vertical temperature-pressure profile (uniform around the globe) equal to the analytic global average profile for the chosen flux conditions and our double-gray absorption coefficients \citep[see Table \ref{tab:params} and][]{Guillot2010}; these coefficients match those used in \citet{Rauscher2017}, chosen to roughly reproduce a profile for a hypothetical planet with Jupiter properties, but 0.1 AU from its host star \citep{Fortney2007}. We ran each simulation for 750 orbits. This length was chosen by extending the zero-obliquity, Jupiter-rotation-rate model out to 3000 orbits and calculating when in the run the emitted fluxes from across the planet were within 1\% of the value at 3000 orbits, indicating that the observable atmosphere had settled into a steady state.

\subsection{Model Results}

\begin{figure}
    \centering
    \includegraphics[width=0.5\textwidth]{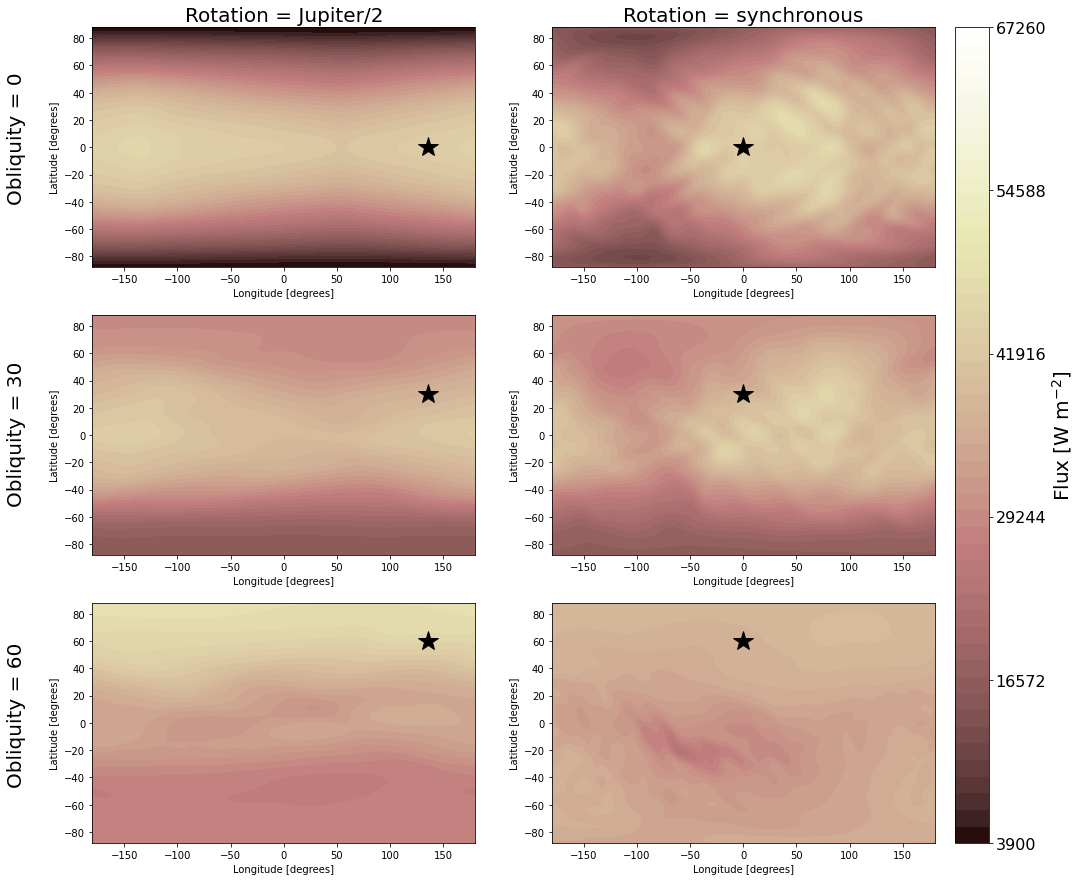}
    \caption{A comparison of the emitted flux patterns from models with different combinations of rotation speed and obliquity. This solstice frame from the animated figure (available in the HTML version of the article, with a duration of 9 seconds) shows the flux map, in the frame rotating with the planet, for models with half the rotation rate of Jupiter (left column) and in synchronous rotation states (right column) and obliquities of 0\degrees, 30\degrees, and 60\degrees~(from top to bottom rows). The star symbol marks the location of the substellar point. The animated figure starts at an equinox point and runs for one full orbit of the planet. During that time the emitted flux pattern changes, with the brightest region slightly lagging behind the location of the substellar point, as discussed in more detail in the main text.}
    \label{fig:cylin_fmaps}
\end{figure}

Our suite of models, sampling the full combination of five rotation rates (from synchronous to as fast as Jupiter) and five obliquities (from 0-90\degrees) exhibits significant diversity in atmospheric circulation patterns and temperature structures, influencing how these physical properties would show up in observations. Figure \ref{fig:cylin_fmaps} provides a representative set of the types of circulation patterns seen in our suite of models, including the presence or lack of seasonal variation for models with or without significant obliquity, respectively. The full set of animations showing maps of the bolometric thermal emission from each model is available on Zenodo at \cite{wJmovies}.

We find a difference in dynamical regimes between the planet models that rotate more quickly ($\omega_\mathrm{Jup}$, $\omega_\mathrm{Jup}/2$, and $\omega_\mathrm{Jup}/4$) and more slowly ($\omega_\mathrm{Jup}/12$ and synchronous $\approx \omega_\mathrm{Jup}/24$), in agreement with previous studies of atmospheric regime transitions for warm Jupiters \citep{Showman2015,Ohno2019a}. The slowly rotating models have larger day--night temperature differences while the quickly rotating models have temperature structures that are more homogenized in longitude but with steeper latitudinal gradients. This can be understood from a combination of two factors: 1) the movement of the irradiation pattern in comparison to the radiative timescale of the atmosphere, and 2) whether the Coriolis force is dominant in balancing global pressure gradients or not. If we characterize the overall atmospheric response by a single radiative timescale, then our suite of models ranges from planets rotating slowly enough that the radiatively driven thermal structure can track the spatial irradiation pattern (i.e., a day-night temperature difference) to planets rotating quickly enough that the gas cannot heat and cool quickly enough to keep up with the changing irradiation pattern, but instead it is the rotationally averaged (i.e., diurnal) irradiation pattern that shapes the radiatively driven thermal structure (for zero obliquity this results in a hotter equator and cooler poles). The dynamical force balances across this regime are explained clearly and extensively in \citet{Showman2015}; in short, in the more quickly rotating models the Coriolis force is stronger and is dominant in balancing the large-scale pressure gradients, while in the slowly rotating models advection is the balancing term in the momentum equation. The characteristic horizontal thermal gradients in an atmosphere are maintained by pressure gradients and \citet{Showman2015} showed that for warm Jupiters the difference between these regimes results in the latitudinal temperature gradients in the quickly rotating models being several times larger than the slowly rotating models, especially away from the equator where the Coriolis force weakens. For non-zero obliquities the atmospheric structures become complex as the irradiation pattern moves in latitude, but the more quickly rotating models will still have temperature structures with stronger latitudinal gradients and minimized longitudinal differences than the slowly rotating models.

The left column of Figure \ref{fig:cylin_fmaps} presents results with behavior typical of our quickly rotating models; shown are models with a rotation rate two times slower than Jupiter. For the models with rotation rates 1, 2, and 4 times slower than Jupiter, we find that for obliquities $\leq30$\degrees~the flux pattern is characterized by a brighter equator and cooler poles while for obliquities $\geq60$\degrees~there is a strong summer-winter hemisphere difference, with some lag time as the atmosphere responds to the changing latitudes of instellation. Uniquely, for the model rotating 4 times more slowly than Jupiter and with an obliquity of 90\degrees, globally the equator remains dimmer than the poles throughout the planet's entire orbit. As we use slower rotation rates for the planet, a bright region to the east of the substellar point becomes more prominent, but the latitudinal flux gradient still dominates the planet's overall global pattern. 

The right column of Figure \ref{fig:cylin_fmaps} shows behavior typical of our slowly rotating models, with flux patterns significantly different from the quickly rotating cases. The models shown here are in synchronous rotation states, but the models rotating roughly twice as quickly (12 times slower than Jupiter) show similar large-scale flux gradients. For the low obliquity cases ($\psi \leq 30$\degrees) the dominant flux pattern is a dayside brighter than the nightside, but with a shift of the brightest region to be eastward of the substellar point. In the case of zero obliquity this pattern remains static, albeit with some slight perturbations from small scale atmospheric waves,\footnote{In the most slowly rotating models, especially those with low obliquity, we see atmospheric waves influencing the flux patterns on the planet, but waves are also visually apparent in all but the most quickly rotating models. In the low obliquity models there are waves that appear to be quasi-stationary relative the substellar point, while in all models there are also waves that can be seen to travel around the globe.} while for the non-synchronous slowly rotating model ($\omega_\mathrm{Jup}/12$) this pattern moves with the substellar point, meaning that it moves in longitude along with the dayside hemisphere. For the slowly rotating models with a low but non-zero obliquity ($\psi=$10\degrees~or 30\degrees) this hemispheric, eastward-shifted pattern oscillates slightly up and down in latitude, following the seasonal cycle. 

The slowly rotating, high obliquity models show complex flux patterns, as demonstrated by the bottom right panel of Figure~\ref{fig:cylin_fmaps}. Due to the poles receiving more flux than the equator over an orbital cycle, they are in general hotter and brighter than the equator. However, as the planet moves toward equinox and the equator receives more direct starlight, the dim ring that had extended around the equator now moves with the changing location of the terminator (the ring is 90\degrees~away from the substellar point). Then at equinox this pattern is broken and the day-night flux difference becomes a significant component of the global flux pattern, before lessening in significance as the substellar point continues to move to higher latitudes.

A more extensive dynamical analysis of these circulation states is beyond the scope of this paper, but we point the reader to the thorough shallow-water analysis of warm Jupiters with various rotation rates, obliquities, and eccentricities in \citet{Ohno2019a} and the aforementioned work of \citet{Showman2015}. We find general agreement between our results and the models in \citet{Ohno2019a}, although our range of rotation rates covers intermediate behavior between their regimes where day-night versus diurnal mean forcing is dominant. We also agree that obliquities less than 18\degrees~produce largely time-invariant atmospheric structure (in the diurnal mean forcing regime); however, our 30\degrees~obliquity model shows some seasonal variation, but without the hemispheric swings in temperature seen by \citet{Ohno2019a}. We now move on to use our simulated set of possible warm Jupiter atmospheric states to quantitatively analyze how secondary eclipse measurements can be affected by the diverse possibilities, and remind the reader that a complementary analysis of thermal phase curves was presented in \citet{Ohno2019b}.

\section{Simulated Secondary Eclipses} \label{sec:eclipses}

\begin{figure}
    \centering
    \includegraphics[width=0.45\textwidth]{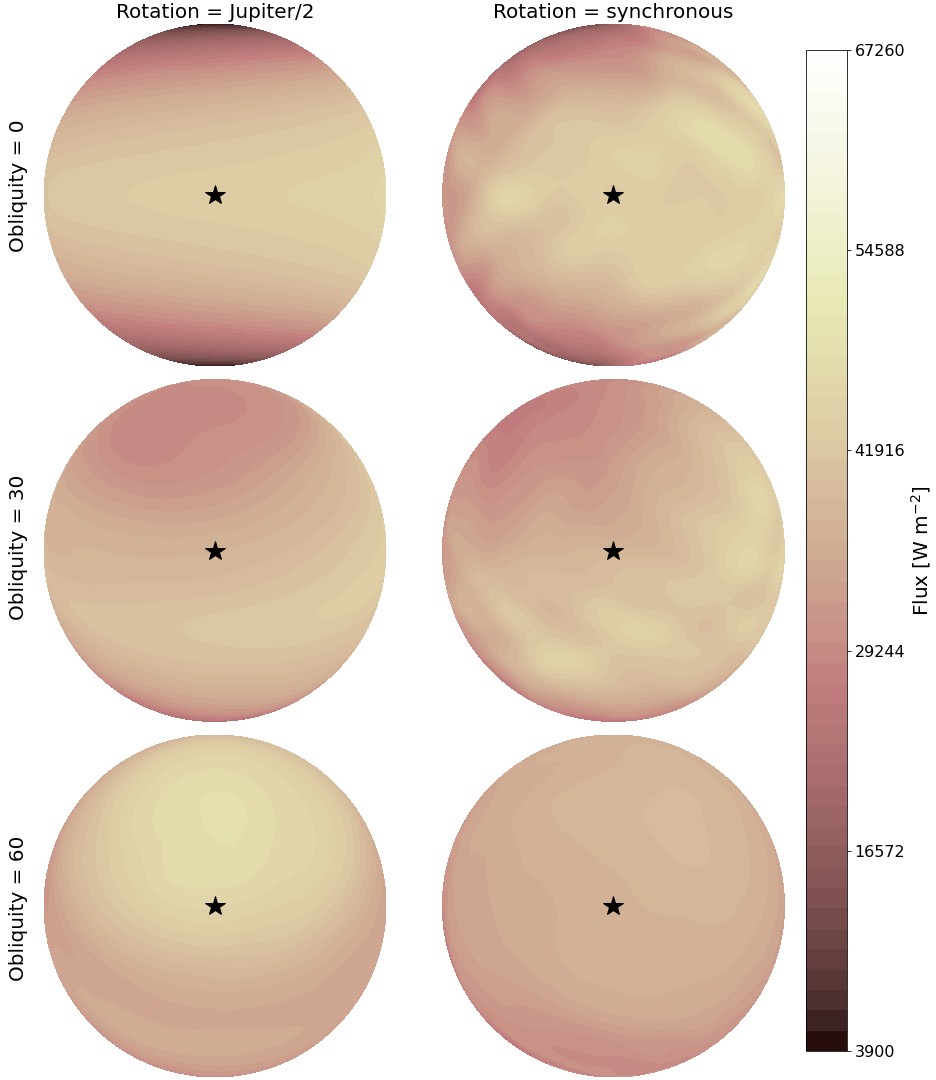}
    \caption{A comparison of the dayside emission patterns from models with different combinations of rotation and obliquity. This frame at solstice from the animated figure (available in the HTML version of the article, with a duration of 9 seconds) shows the same set of flux maps as in Figure \ref{fig:cylin_fmaps}, but shown from the direction of the star, since this is what the dayside of the planet would look like if an observer were oriented at that moment in time to catch the planet in secondary eclipse (for an edge-on orbit, with the stellar limb progressing horizontally across the planet disk). As in Figure \ref{fig:cylin_fmaps}, these are the models with half the rotation rate of Jupiter (left column) and in synchronous rotation states (right column) and obliquities of 0\degrees, 30\degrees, and 60\degrees~(from top to bottom rows) and the star symbol marks the location of the substellar point. The animated figure starts at an equinox point and runs for one full orbit of the planet. Each frame in the animated figure shows one possible dayside flux pattern that we could observe for the secondary eclipse of the planet; depending on our orientation relative to the system, we would only ever observe the planet as in that single frame.}
    \label{fig:ortho_maps}
\end{figure}

From our set of 3D atmospheric models we simulate secondary eclipse light curves using the software package \texttt{starry} \citep{Luger2019}, which takes as input the two-dimensional planet flux maps and calculates simulated light curves from the defined system geometry, including the appropriate tilts and rotation of each model. We use the system parameters in Table \ref{tab:params}, choosing to assume an orbital inclination of 88.5\degrees, which results in an impact parameter of $\sim$0.5. For this, or any other transiting planet with a non-zero obliquity, we may happen to observe secondary eclipse at any point during its seasonal cycle, and we will always observe the planet at that point every eclipse. For example, the secondary eclipse of the planet may always occur near equinox (in which case the rotation vector of the planet must be tilted near-parallel to our line of sight) or always near a solstice (in which case the rotation vector will be tilted toward or away from us). Using 90 snapshots of the flux from the planet, evenly spaced throughout its last simulated orbit and so evenly sampling its seasonal cycle, we orient each map to the correct geometry for secondary eclipse \citep[as shown in Figure \ref{fig:ortho_maps}, with the full set of animations available on Zenodo at][]{wJmovies} and calculate a simulated light curve. We include an equal time before and after eclipse as during it, to mimic a typical eclipse observation, resulting in a simulated observation lasting 8.64 hours.

\begin{figure*}[ht]
    \centering
    \includegraphics[width=0.9\textwidth]{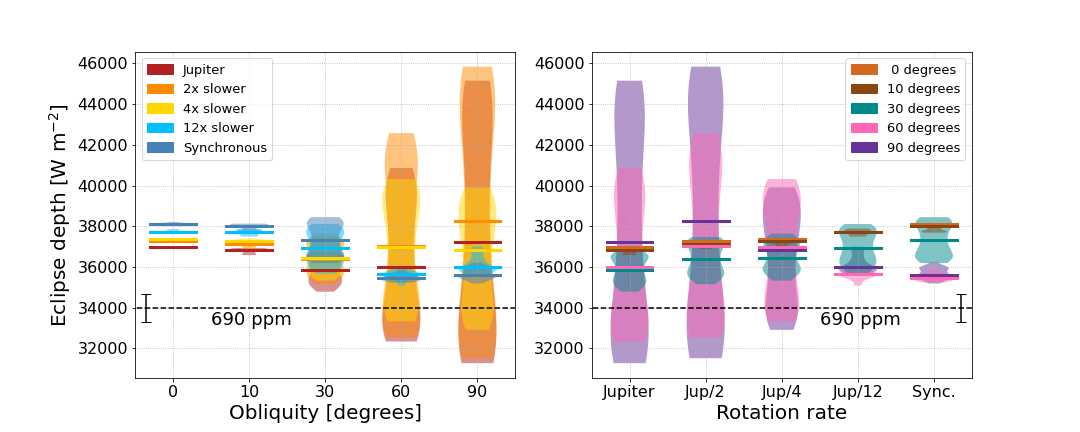}
    \caption{The eclipse depths for our hypothetical warm Jupiter, from models with different combinations of rotation rate and obliquity. For each model the eclipse depth depends on the timing of the eclipse relative to the planet's seasonal cycle. The median value for each model is marked as a line, with the distribution of possible values indicated by the violin plots. Marked as a dashed line is the eclipse depth this planet would have if it had a uniform temperature equal to its zero-albedo equilibrium temperature, from which we can also estimate the depth relative to the stellar flux (in ppm: parts per million) at 10 micron and an early evaluation of the precision of the MIRI instrument on JWST \citep[from][]{Bouwman2023} with the error bars. For most rotation rates a non-zero obliquity can cause deviations from the zero obliquity expectation, ranging from insignificant differences for $\psi \leq 10$\degrees, to a 5\% effect at $\psi=30$\degrees~(potentially differentiated by JWST), to a 15-20\% effect at $\psi \geq 60$\degrees~(easily differentiated by JWST). Eclipse depth alone cannot constrain a planet's rotation rate and obliquity, but if those parameters are otherwise unconstrained, they can induce uncertainty in the interpretation of a measured eclipse depth.}
    \label{fig:edepth}
\end{figure*}

The first-order quantity measured at secondary eclipse is the integrated flux from the dayside of the planet, which sets the depth of the eclipse. Figure \ref{fig:edepth} shows these eclipse depths, both the median values and full ranges for each model of the planet, as a function of the assumed rotation rate and obliquity. Immediately apparent is that a wide range of eclipse depths are possible and planets with larger obliquities and faster rotation rates exhibit a greater variation in possibilities, depending on when secondary eclipse is observed relative to the seasonal cycle. This can be understood in terms of the dynamical states found above: only the quickly rotating models are able to maintain strong temperature gradients in latitude. For low obliquity cases, this results in the equator being warmer than the poles, with little variation throughout the planet's annual cycle; however, at high obliquity strong seasonal changes result in different flux patterns being observed for eclipses at different times throughout the orbit. If the secondary eclipse happens around the time of a solstice, we will see a very hot, bright dayside, whereas if secondary eclipse occurs around an equinox, we will instead see some of each hemisphere, both of which will be at only moderate temperatures. Meanwhile, the slowly rotating models have large day-night temperature differences and little-to-no seasonal variation (at low obliquity) or more globally homogenized temperatures (at high obliquity), which effectively also diminishes any seasonal variation. Since we always observe the dayside hemisphere in eclipse, this results in larger eclipse depths at low obliquity and somewhat smaller eclipse depths at high obliquity, but a small range of possible values for each state. At low obliquities ($\psi \le 30$\degrees) and/or slow rotation (synchronous or $\omega_J/12$) the eclipse depth variation is smaller ($\le 5$\%), while higher obliquity, quickly rotating models have a much larger range of values, up to $\sim$35\%.  

Note that in almost all cases the eclipse depths are greater than the value for a planet with uniform temperature (equal to the zero-albedo equilibrium temperature), which can generally be interpreted as some inefficiency in moving heat from the day to night side; however, there are some instances (at high obliquity and fast rotation) that the eclipse depth would be below the equilibrium temperature. One example of why this happens can be understood for the slightly odd case of our model with a rotation rate of $\omega_\mathrm{Jup}/4$ and an obliquity of 90\degrees. Throughout this planet's orbit the equator is always colder than the poles. This means that for viewing orientations where the eclipse happens near equinox, we are preferentially seeing the cooler and dimmer regions of the planet, bringing the hemispheric flux below that of an equivalent uniform temperature planet. Traditionally this would be interpreted as a non-zero albedo on the planet, resulting in an overall cooler atmospheric state, but here that would be incorrect. We note this as a point of caution for the interpretation of warm Jupiter eclipse depths: seasonal effects can complicate the relation between albedo, day-night heat redistribution, and dayside flux \citep{Cowan2011b}.

We estimate the overall eclipse depth signal for our hypothetical planet by calculating the blackbody planet-to-star flux ratio: 
\begin{equation}
F_p/F_s=(R_p/R_s)^2(B_\lambda(T_p)/B_\lambda(T_s)),     \label{eqn:fpfs}
\end{equation}
where $B_\lambda(T)$ is the Planck function. Using the parameters for our system and assuming the planet's temperature equals its equilibrium temperature, we find a secondary eclipse depth of 690 ppm (parts-per-million) at 10 micron.\footnote{Commensurate with the other simplifications in this analysis, we choose to estimate flux ratios (here and later) at a single wavelength instead of integrated across the MIRI wavelength range. The planet-to-star flux ratio for this set up is a factor of $\sim$2 smaller at 5 micron than at 10 micron, so our single-wavelength values may be slight overestimates of what a more detailed calculation would find.} This is easily within the capabilities of the MIRI instrument on JWST. As an early example, \citet{Bouwman2023} demonstrated that MIRI is able to measure the transit of a super-Earth to a precision of 13-15 ppm in the band averaged transit depth. Thus JWST would have no difficulty in measuring the secondary eclipse of our warm Jupiter, but also would be able to differentiate between the range of predictions offered by our models, as 14 ppm translates into about a 700 W~m$^{-2}$ precision on the eclipse depth for this system. In other words, as JWST begins to observe secondary eclipses of warm Jupiters, beyond the regime where their rotation vectors are synchronized and aligned with their orbit, \textit{we should expect to see an increase in the dispersion of measured eclipse depths---if nature produces a range of rotation rates and obliquities among this population}.

\section{Potentially Constraining Rotation Vectors with Eclipse Mapping} \label{sec:mapping}

If measured at sufficiently high precision, the secondary eclipse observation of an exoplanet can also constrain the latitude-longitude brightness map on the planet's dayside through the eclipse mapping technique \citep{Williams2006,Rauscher2007b,Agol2010,deWit2012,Majeau2012,Dobbs-Dixon2015}. Since the dayside flux pattern on our hypothetical warm Jupiter is strongly shaped by its assumed rotation rate and obliquity, here we assess whether there are spatial patterns detectable in eclipse mapping that could be used to uniquely constrain the planet's rotation vector. There are two pieces to this question: 1) what spatial information is intrinsically encoded in secondary eclipse observations, and 2) whether those signals are large enough to measure, compared to instrumental precision. We address each of these in turn.

In order to identify what spatial structures in our planet models may be accessible to eclipse mapping, we fit each of the simulated secondary eclipse curves, described above, with the lightcurve for an uniform planet and a set of ``eigencurves". These orthogonal curves are mathematically optimized and sorted such that the maximum spatial information can be extracted from the light curves \citep[using the method of][and we refer the reader to that reference for more details]{Rauscher2018}. We use the \texttt{scipy curve\_fit} routine and do not include any noise, as the goal of this fit is to identify contributions from possible spatial inhomogeneity, not to assess whether it is measurable (at this stage). There are additional complications in eclipse mapping when the rotation vector of the planet is unknown, as this complicates the translation from observed flux at some time to which regions on the planet are visible. \citet{Adams2023} explored this complexity and determined that the mapping of the planet should be able to correctly infer large-scale features---regardless of intrinsic obliquity and viewing angle---as long as the planet's rotation period remains longer than the timescale of ingress and egress, which is indeed the regime we consider.

We use ten eigencurves to fit to the simulated eclipse curves, allowing a large number of components in order to capture small-scale signals in the light curves (associated with small features in the maps) but limited to ten because higher order terms have eigenvalues more than two orders of magnitude smaller, implying much less potential contribution to the signal. This resulted in a set of ten fit eigencurve coefficients for each rotation rate/obliquity model, for each of the 90 possible secondary eclipse curves simulated throughout its yearly cycle. We examined the numerical distributions of each of these ten coefficients, for combinations of coefficient values that could point toward specific underlying rotation rate/obliquity states. 

\begin{figure*}
    \centering
    \includegraphics[width=0.475\textwidth]{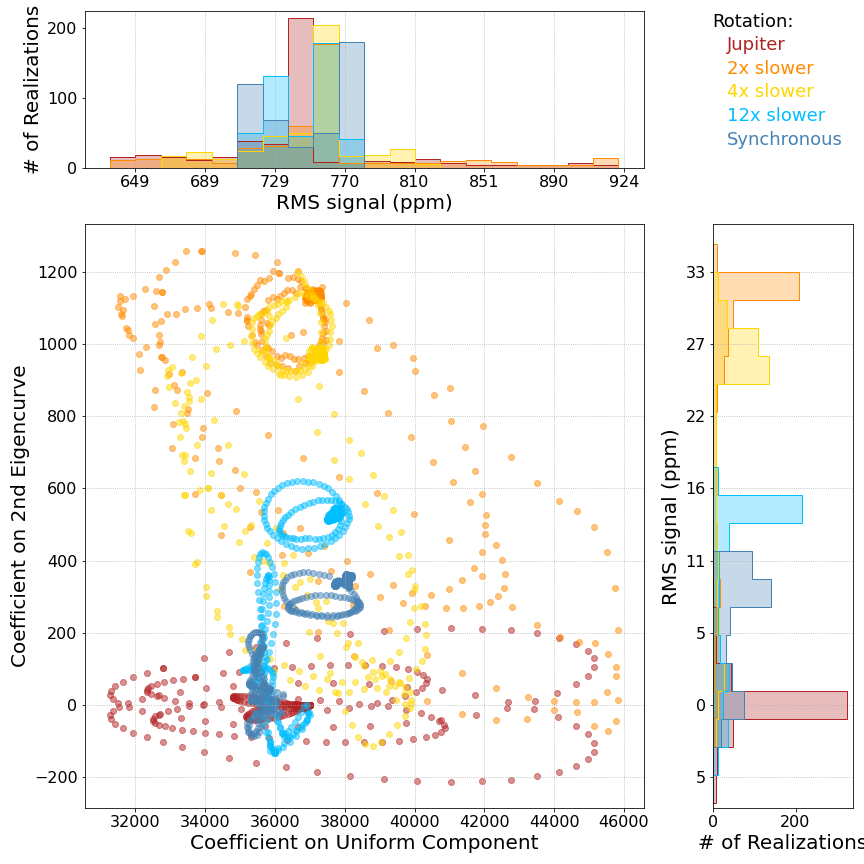}
    \includegraphics[width=0.475\textwidth]{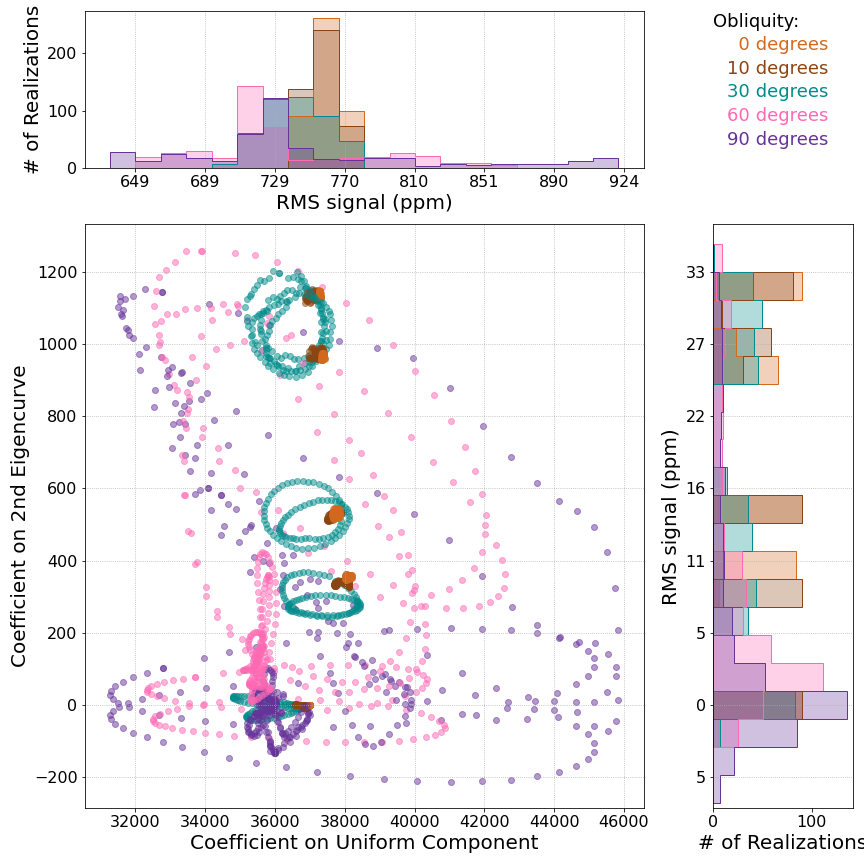}
    \caption{Plots of the coefficients for the uniform disk and second eigencurve components, found from fitting to the 90 simulated secondary eclipse observations for each of the 25 rotation rate/obliquity models. The color of the point indicates the rotation rate \textit{(left)} or obliquity \textit{(right)} of the model. The coefficients indicate the strength of that signal in the observation; on the axes for the histograms we translate this into an RMS signal in parts-per-million, as described in the text. The spatial brightness structure of the planet's dayside, if it could be measured by eclipse mapping, would constrain the planet's rotation state. }
    \label{fig:mapping}
\end{figure*}

In Figure \ref{fig:mapping} we show an example of how eclipse mapping could be used to try to constrain the warm Jupiter's rotation vector. Of the ten eigencurves included in the fit, the second eigencurve is generally the largest signal in the simulated eclipse curves, after the uniform component. This is because this eigencurve is associated with an eigenmap that captures any large-scale east-west flux gradient and so also induces a large ``phase curve" signal outside of eclipse. Conveniently, as shown in Figure \ref{fig:mapping}, the strength of any east-west flux gradient on the planet shows strong dependence on the planet's rotation rate. Although not fully unique, this signal in a planet's secondary eclipse curve could help to differentiate between possible rotation rates on the planet, due to the changes in circulation state discussed above. Used in combination with the eclipse depth, it could perhaps also be possible to isolate possible obliquity states as well, as the models with different rotation rates and obliquities sometimes occupy non-overlapping regions in the parameter space of these two observed properties.

In addition, we calculated the signal that each of these eigencurves would contribute to the overall secondary eclipse measurement, as parts-per-million in the planet-to-star flux ratio. Most of the eigencurves are only significantly non-zero during the times of ingress and egress, and since they relate to spatial patterns that increase or decrease regions of the planet's brightness map away from uniform, they can have positive or negative values. Thus, in order to quantify how much each component contributed to the overall secondary eclipse measurement, we calculated a normalized root-mean-squared (RMS) value: signal$=\sqrt{\int F_e^2 dt/\int F_0^2 dt}$, where the eigencurve ($F_e$) and uniform disk eclipse curve ($F_0$) are integrated over the entire time of the simulated observation. We then use Equation \ref{eqn:fpfs} again to estimate the expected signal at 10 micron.

In Figure \ref{fig:mapping} we can see that, while the overall eclipse depth should be easily measurable by JWST, the signal from any east-west brightness structure is comparable to the precision of JWST/MIRI \citep{Bouwman2023}. This second eigencurve had the largest signal of any of the eigencurves due to its partial phase curve component; others ranged in contribution from $\sim$1-15 ppm. These signals are tantalizingly at the edge of JWST capabilities. 

\section{Discussion} \label{sec:discussion}

There are a few caveats to consider in this work. For one, we have chosen a 10-day orbital period for our hypothetical warm Jupiter, which for most systems would place it within the regime where tidal synchronization and alignment should be expected, unless it was a younger-than-average system.\footnote{It is also interesting to note that for a warm Jupiter on a ten day orbit but with a substantial obliquity, we might expect active tidal dissipation to produce a significant internal heat flux in the planet, potentially inflating the planet's radius \citep[e.g.,][]{Millholland2019c} and influencing the atmospheric dynamics \citep[e.g.,][]{Komacek2022}. We neglect these considerations here.} This choice also results in stronger stellar forcing and likely larger spatial differences than warm Jupiters on longer orbital periods, representing an observationally optimistic set-up. We also considered a solar twin for the host star, but several important factors depend on the stellar host, all else being equal: the efficiency of tidal synchronization and alignment, the planet-to-star flux ratio, and the expected strength of atmospheric seasonal responses. \citet{Tan2022} showed that when a seasonal cycle induces changes in stellar forcing, temporal variations in the planet's atmospheric response depend to first order on the ratio between its orbital period and radiative timescale, which for Jupiter-like atmosphere conditions is:
\begin{equation}
    \frac{P_{\mathrm{orb}}}{\tau_{\mathrm{rad}}} \approx 5.94 \left( \frac{T_s}{T_\odot} \right)^3  \left( \frac{R_s}{R_\odot} \right)^{3/2}  \left( \frac{M_s}{M_\odot} \right)^{-1/2}. \label{eqn:season_timescales}
\end{equation}
For stellar masses $\lesssim 0.6 M_{\odot}$ the radiative timescale exceeds the orbital timescale and seasonal effects become muted.

\begin{figure*}
    \centering
    \includegraphics[width=0.95\textwidth]{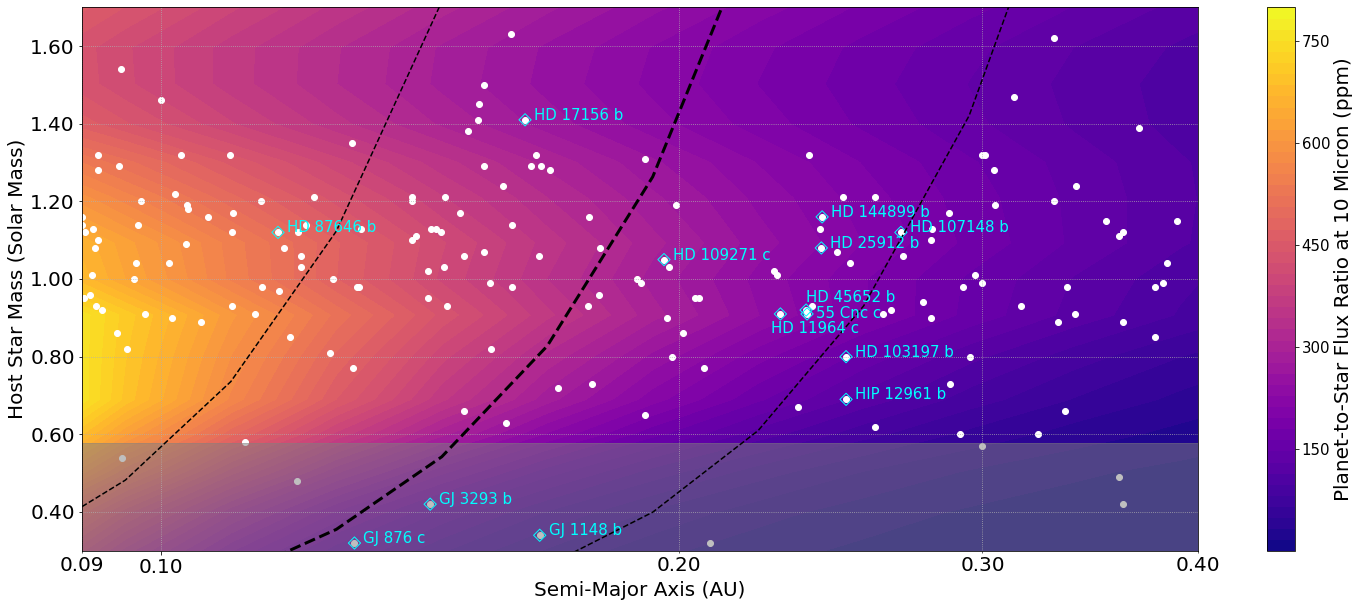}
    \caption{To assess the feasibility of measuring seasonal effects on gaseous exoplanets, we compare the parameter space over which tidal forces from the star should have or have not synchronized and aligned a planet's rotation vector to the regime where a planet's atmospheric response to a seasonal cycle may be minimal and to expected planet-to-star flux ratios and two different wavelengths. The middle black dashed line marks the boundary where the tidal timescale for a Jupiter analog equals an assumed system age of 4.14 Gyr, so planets to the left of this should be in synchronous and aligned rotation states, while the lines to the left and right of that are for factor of ten differences in the assumed system age, tidal $Q$ parameter, or initial rotation rate (Equation \ref{eqn:tidal}). The gray region with stellar masses $\lesssim 0.6 M_{\odot}$ is where \citet{Tan2022} identified weak atmospheric responses to seasonal cycles. The filled color contours show predicted planet-to-star flux ratios (in parts per million) at 10 micron. The white dots represent known gaseous exoplanets within this parameter space, with labels given to those particularly bright planets ($K \leq 8.131$) whose estimated tidal timescales are comparable to their system ages ($0.25 <$ age/$\tau_{\mathrm{tidal}}<2$). The best targets to search for seasonal effects are likely planets at or below a Solar mass.}    \label{fig:stellarmass}
\end{figure*}

We compare the influence of these various effects for Jupiter-like planets orbiting stars of different masses for a range of semi-major axes in Figure \ref{fig:stellarmass}. To calculate the planet-to-star flux ratio we use Equation \ref{eqn:fpfs}, assuming Jupiter's radius for all planets and adjusting the planet temperature to be the zero-albedo equilibrium temperature at each semi-major axis. For the stellar radii and temperatures as a function of stellar mass, we use empirical averages for mass bins across the main sequence \citep[from Table 6 of][]{Eker2018}. These stellar parameters are also used in Equation \ref{eqn:season_timescales}, obtaining agreement on the stellar mass limit from \citet{Tan2022}. We use the original form of Equation \ref{eqn:tidal} from \citet{Guillot1996}, written as a function of semi-major axis instead of orbital period, and assume the same planet parameters as in Section \ref{sec:vector_expectations}. Figure \ref{fig:stellarmass} paints a complex picture; for a given semi-major axis, planets around lower mass stars are less likely to be tidally aligned and synchronized, but their atmospheres should have weaker seasonal responses. In terms of observability, the planet-to-star flux ratio at 10 micron peaks slightly below one Solar mass. At this wavelength we are not always within the Rayleigh-Jeans tail of the Planck function for these objects and so the trend with stellar mass is complex. This demonstrates that it is not trivial to identify the best targets for observing atmospheric seasonal responses, but in general we may prefer near-Solar stellar hosts.

We used double-gray radiative transfer within our GCM; including more realistic radiative transfer would alter the temperature-pressure profiles and radiative forcing of the dynamics, but a study in the hot Jupiter context has shown that double-gray radiative transfer captures the same large-scale qualitative features as more complex models \citep{Lee2021}, giving some confidence that our warm Jupiter results would similarly not be too strongly changed. We have also neglected other complicating atmospheric physics, such as disequilibrium chemistry and clouds. Both effects have the potential to alter the radiative forcing of the atmosphere and resulting dynamics, but would have to be unexpectedly strong in order to significantly change the circulation patterns \citep[e.g., when half of a hot Jupiter's dayside was covered with highly reflective clouds in][this did not disrupt the standard hot Jupiter circulation pattern]{Roman2017}. Both clouds and disequilibrium chemistry, in addition to considering spectral emission from the planet instead of bolometric flux, could either diminish or enhance the flux differences we predict across the planet, depending on whether there is global cloud coverage or only cloud formation in the coolest regions \citep[e.g.,][]{Roman2021}, respectively, and how disequilibrium chemistry changed the distribution of species' abundances across the globe. Spectral variation in the emission from regions of the planets with different temperatures, chemical abundances, and/or clouds could potentially provide more information about the planet's circulation state, through spectral eclipse mapping \citep{Mansfield2020,Challener2022}, but that would require even more precise measurements in order to constrain wavelength-dependent spatial patterns.

Finally, while we have focused on the spatial information available from eclipse mapping, one of our main findings is that the largest non-uniform spatial signal comes from the eigencurve that includes partial phase curve information. Since we also found that the corresponding spatial pattern---an east-west asymmetry in the brightness distribution---is related to the planet's rotation rate, it may be that the partial phase curve information available in JWST secondary eclipse observations of warm Jupiters will be enough to begin to differentiate their rotation states. However, this does rely on being able to accurately correct out any instrumental systematics that could be confused with the phase curve signal. Work by \citet{Schlawin2023} finds that the east-west structure can be particularly degenerate with systematic signals, but that statistical evidence such as the Bayesian Information Criterion may be able to differentiate between eigencurve fits that accurately capture the systematics and planet signal from those that do not. There may similarly be clues that the systematics are being incorrectly fit if the planet map contains nonphysical structures.

\section{Conclusions} \label{sec:conclusions}

As we make atmospheric characterization measurements of planets on longer orbital periods, we will move into a regime where we can no longer assume that a planet's rotation period and axis are equal to and aligned with its orbital period and axis, as we do for hot Jupiters. From simple timescale estimates, we may expect that warm Jupiters with orbital periods $\gtrsim$20 days have \textit{a priori} unknown rotation vectors. 

We ran three-dimensional atmospheric circulation models of a hypothetical warm Jupiter, over a range of rotation rates and axial obliquities. We found moderate and strong seasonal variations in the atmospheres of planets with obliquities of 30\degrees~and $>$60\degrees, respectively. At low obliquities ($\leq$30\degrees) the dominant atmospheric structures depend on the rotation rate, with more quickly rotating models (from 1-4 times slower than Jupiter) showing brighter equators than poles, while the more slowly rotating models (at or about twice the synchronous rotation rate) have an eastward-shifted hot-spot in addition to the latitudinal gradient.

To examine the observable implications for atmospheric characterization of non-tidally synchronized and aligned planets, we calculated simulated secondary eclipse spectra from our models, for 90 points throughout each orbit to cover the full seasonal cycle. We showed that the secondary eclipse depth is a strong function of rotation rate, obliquity, and viewing orientation. For the higher obliquity and more quickly rotating models, where in the seasonal cycle we happen to catch the planet in secondary eclipse strongly influences how bright the dayside will be, with a spread in possible values as large as 35\%. This can complicate the interpretation of secondary eclipse depths, especially for the outlier cases when the eclipse depth is lower than that expected for an uniform planet at the equilibrium temperature. The dispersion in possible secondary eclipse depths is large enough for JWST to be able to measure, resulting in our main prediction: if nature produces warm Jupiters with a range of rotation rates and obliquities, as a consequence we should expect to measure an increase in the dispersion of secondary eclipse depths for the population of gaseous planets beyond the tidal boundary.

We evaluate whether eclipse mapping could be used to measure the brightness distribution across the planet and thereby constrain its rotation vector, by calculating ``eigencurves" and fitting to the simulated eclipse curves. We find that there are unique spatial components that, in combination, could be used to constrain the rotation rate and obliquity, but that the expected signals are comparable to or below current estimates of JWST precision. Further work may need to be done, both in modeling and JWST data analysis, to determine whether warm Jupiters beyond the tidal boundary can be eclipse mapped by JWST.

As JWST characterization efforts push out to measure secondary eclipses of longer orbital period planets, we will learn about atmospheric physics in new and interesting regimes. Here we have shown that once we move beyond the tidal synchronization and alignment boundary, any diversity in the rotation vectors of warm Jupiters should be observable as a population-level dispersion in their secondary eclipse depths. While eclipse mapping could potentially constrain a planet's rotation rate and obliquity, we found that the eclipse mapping signals are near or below the initial precision JWST/MIRI seems able to achieve, for our hypothetical warm Jupiter on an optimistically bright 10-day orbit. This implies that most warm Jupiters beyond the tidal boundary may produce smaller signals, but perhaps as analysis of JWST data continues to mature we will discover that these minute signatures are detectable. In that happy circumstance, eclipse mapping of warm Jupiters could be used to constrain the rotation vectors of these planets, which in turn would help to inform theories for their formation and evolution.

\begin{acknowledgments}
We thank the anonymous referee, whose comments helped to improve and clarify this manuscript. This research was supported by NASA Astrophysics Theory Program grant NNX17AG25G and made use of the NASA Exoplanet Archive, which is operated by the California Institute of Technology, under contract with the National Aeronautics and Space Administration under the Exoplanet Exploration Program.
\end{acknowledgments}


\software{
          Jupyter notebook \citep{jupyter},
          NumPy \citep{numpy},
          SciPy \citep{SciPy}, 
          starry \citep{Luger2019}
          }




\bibliography{biblio}{}
\bibliographystyle{aasjournal}

\end{document}